\documentclass[final,3p,12pt,times]{elsarticle}

\usepackage[utf8]{inputenc}

\newcommand{\bes}{\begin{eqnarray*}}
\newcommand{\ees}{\end{eqnarray*}}

\newcommand{\bel}[1]{\begin{eqnarray}\label{#1}}
\newcommand{\be}{\begin{eqnarray}}
\newcommand{\ee}{\end{eqnarray}}

\newcommand{\rf}[1]{Eq.~(\ref{#1})}
\newcommand{\rfn}[1]{~(\ref{#1})}

\newcommand{\rff}[1]{Fig.~\ref{#1}}
\newcommand{\rfc}[1]{Ref.~\cite{#1}}

\newcommand{\ackno}{\section*{Acknowledgements}}

\newcommand{\f}[2]{\frac{#1}{#2}}

\newcommand{\sym}{${\mathcal N}=4$}
\newcommand{\symm}{${\mathcal N}=4$ SYM}

\newcommand{\ed}{{\cal E}}       
\newcommand{\peq}{{\cal P}}     
\newcommand{\pT}{{\cal P}_T}   
\newcommand{\pL}{{\cal P}_L}   

\newcommand{\tpi}{\tau_{\pi}}

\newcommand{\bort}{\mathcal{B}}
\newcommand{\borta}{\tilde{\mathcal{B}}}
\newcommand{\ibort}{\mathcal{S}}
\newcommand{\pa}{{\cal A}}
\newcommand{\pac}{{a}}

\newcommand{\mysection}[1]{\section{{#1}}}

\journal{Physics Letters B}

\begin{document}

\begin{frontmatter}

\title{On the hydrodynamic attractor of Yang-Mills plasma}

\author[1,2]{Micha\l\ Spali\'nski}
\address[1]{Physics Department, University of Bia{\l}ystok,
  Konstantego Cio\l kowskiego 1L,
  15-245 Bia\l ystok, Poland}
\address[2]{National Centre for Nuclear Research,
  Ho{\.z}a 69,
  00-681 Warsaw, Poland}

\begin{abstract}

There is mounting evidence suggesting that relativistic hydrodynamics becomes relevant for the physics of
quark-gluon plasma as the result of nonhydrodynamic modes decaying to an attractor apparent even when
the system is far from local equilibrium. Here we determine this attractor for Bjorken flow in \sym\
supersymmetric Yang-Mills theory (SYM) using Borel summation of the gradient expansion of the expectation
value of the energy momentum tensor. By comparing the result to numerical simulations of the flow based on
the AdS/CFT correspondence we show that it provides an accurate and unambiguous approximation of
the hydrodynamic attractor in this system. This development has important implications for the formulation of
effective theories of hydrodynamics.

\end{abstract}


\end{frontmatter}

\mysection{Introduction} Heavy-ion collision experiments and their phenomenological description have lead to the
realization that relativistic hydrodynamics works very well rather far outside its traditionally understood domain of
validity. Variants of M\"uller-Israel-Stewart (MIS) theory~\cite{Muller:1967zza,Israel:1976tn,Israel:1979wp} have
successfully been applied in rather extreme conditions, which could hardly be assumed to be close to local equilibrium.
Furthermore, model calculations exist where it is possible to study the emergence of universal, hydrodynamic behaviour
and test to what extent an effective description in terms of hydrodynamics can match microscopic
results~\cite{Florkowski:2017olj}. Such calculations were initially carried out in \symm\ using the AdS/CFT
correspondence~\cite{Chesler:2008hg,Heller:2011ju,Jankowski:2014lna}, but
similar studies have since also been performed in models of kinetic
theory~\cite{Keegan:2015avk,Kurkela:2015qoa,Heller:2016rtz}. The conclusion from these investigations is that the domain
of validity of a hydrodynamic description is delimited by the decay of nonhydrodynamic
modes~\cite{Chesler:2008hg,Heller:2011ju,Spalinski:2016fnj,Romatschke:2016hle}. The outcome of this transition to
hydrodynamics (``hydronization'') is that the system reaches a hydrodynamic attractor~\cite{Heller:2015dha} which
governs its subsequent evolution toward equilibrium. This attractor is a special solution to which generic histories
decay exponentially, and do so well before local equilibrium sets in. It incorporates all orders of the hydrodynamic
gradient expansion, and at sufficiently late times coincides with the predictions of relativistic Navier-Stokes theory.
The existence of an attractor in this sense is a critically important issue for hydrodynamics, because it defines its
very meaning. It has conceptual as well as practical implications for the formulation of hydrodynamic theories in
general as well as for their application to the physics of quark-gluon plasma.
%

%
Attractor behaviour was first identified explicitly in the differential
equations of hydrodynamics~\cite{Heller:2015dha,Aniceto:2015mto}.  An
outstanding problem is the determination of such attractors at the microscopic
level~\cite{Romatschke:2017vte,Florkowski:2017olj}. The first calculations of
this type were described by Romatschke~\cite{Romatschke:2017vte}, who found
approximate attractor solutions in the context of kinetic theory and \symm\ by
scanning for the corresponding initial conditions.  The purpose of this Letter
is to argue that the Borel sum of the hydrodynamic gradient expansion provides
a direct way of estimating the attractor. While at late times this calculation
clearly must give the correct result (which coincides with the prediction of
Navier-Stokes hydrodynamics) it is not obvious {\em a priori} that this
calculation gives an accurate estimate at earlier times.  We will however show
explicitly that the result of Borel summation does indeed act as an attractor
for histories of Bjorken flow simulated using techniques based on the AdS/CFT
correspondence. This should be viewed in the context of the idea that higher
orders of the gradient expansion may be relevant for real-world
physics~\cite{Lublinsky:2007mm,Lublinsky:2011cw,Heller:2013fn,Bu:2014sia}.

An important point is that the hydrodynamic gradient expansion is the leading
element of a transseries~\cite{Heller:2015dha}, and in general the higher
order elements (``instanton sectors'') play an important role in defining the
summation properly. These transseries sectors involve integration constants
which need to be fixed. However, their contributions are exponentially
suppressed and it is tempting to ignore them as a first approximation. Such an
approach will definitely fail at sufficiently early times (before the
exponential suppression sets in). However, we will see that it works fine for
$\tau T>0.3$, and this is enough to see that the result of the Borel sum acts
as an attractor well before the Navier-Stokes approximation to hydrodynamics
becomes accurate at $\tau T \approx 0.7$~\cite{Jankowski:2014lna}.

A critical issue for Borel summation is the location of singularities of the
analytic continuation of the Borel transform. These singularities reflect the
spectrum of nonhydrodynamic modes -- both at the microscopic
level~\cite{Heller:2013fn} and in
hydrodynamics~\cite{Heller:2015dha,Aniceto:2015mto}. An important testing
ground for the feasibility and robustness of Borel summation of the gradient
series of \symm\ is the hydrodynamic theory proposed in~\cite{Heller:2014wfa},
which we will refer to as HJSW. This theory extends Navier-Stokes
hydrodynamics by adding degrees of freedom which mimic the least-damped
nonhydrodynamic modes of \symm\ plasma (known from calculations of quasinormal
modes of black branes~\cite{Kovtun:2005ev}). This results in the same leading
singularities~\cite{Aniceto:2015mto} as those identified at the microscopic
level in \rfc{Heller:2013fn}.  This should be contrasted with BRSSS
hydrodynamics~\cite{Baier:2007ix}, which instead involves only purely decaying
modes.

In the case of BRSSS theory one cannot ignore the transseries sectors even as
an approximation, because the analytically-continued Borel transform of the
hydrodynamic series has branch-point singularities on the real axis
(reflecting the purely-decaying MIS nonhydrodynamic mode) and this leads to a
complex summation ambiguity. The addition of transseries sectors (which are
constrained by resurgence
relations~\cite{Heller:2015dha,Aniceto:2015mto,Basar:2015ava}) resolves this
ambiguity, but requires an integration constant (the transseries parameter) to
be set correctly by comparing the result of the summation to the numerical
calculation of the attractor. Luckily, this issue does not arise in \symm, nor
in HJSW hydrodynamics, because in these cases singularities of the analytic
continuation of the Borel transform occur off the real axis. Thus, omitting
the instanton sectors is a reasonable first approximation, which is what we
focus on here.

As a way of determining the range of proper-time where the Borel sum can be
expected to give an accurate estimate of the attractor we first calculate the
Borel sum of the gradient expansion in the case of HJSW hydrodynamics, where
it is easy to check the validity of the answer. The result is unique,
unambiguous, and coincides (even at rather early times) with the attractor
determined directly from the hydrodynamic equations. This sets the stage for
the main theme of this Letter: the Borel summation of the gradient series of
\symm. This is technically no more challenging than the calculation for HJSW
theory, but its significance is that it provides an example of a hydrodynamic
attractor obtained directly from a microscopic calculation. This result can
only be fully appreciated by inspecting the behaviour of numerically simulated
histories of boost-invariant expansion in \symm.  A very important point to
note is that while the attractor coincides with first order hydrodynamics at
late times, it turns out to be quite distinct from it even at moderate
times. This has implications of foundational nature for relativistic
hydrodynamics. A fuller discussion of this result and its ramifications can be
found in the concluding section.

\mysection{Bjorken flow} Throughout this paper we work with Bjorken flow \cite{Bjorken:1982qr}, which
imposes powerful simplifying symmetry constraints. We use proper time -- rapidity coordinates $\tau, Y$
related to Minkowski lab-frame coordinates $t, z$ by $t = \tau \cosh Y$ and $z = \tau \sinh Y$ where $z$ is
aligned along the collision axis. A system undergoing Bjorken flow has eigenvalues of the expectation value of
the energy momentum tensor
\bel{Tab}
T^{\mu \nu} = \mathrm{diag}(\ed, \pL, \pT, \pT)^{\mu\nu}
\ee
which are functions of the proper time $\tau$ alone. In a conformal theory, the conditions of
tracelessness and conservation can be expressed as~\cite{Janik:2005zt}
\bel{PLT}
\pL = - \ed - \tau\, \dot{\ed}\ , \quad
\pT =  \ed + \f{1}{2} \tau\, \dot{\ed}.
\ee
The departure of these quantities from the equilibrium pressure at the same
energy density, $\peq \equiv \ed/3$, is a measure of how far a given state is
from local equilibrium. This is conveniently captured by
the pressure anisotropy
\bel{rdef}
\pa \equiv \f{\pT-\pL}{\peq}
\ee
which we will study as a function not of the proper time $\tau$, but of the dimensionless ``clock variable'' $w\equiv
T\,\tau$, where $T$ is the effective temperature (defined as the temperature of the equilibrium state with the same
energy density). It is critically important to compare states of the system at different values of this dimensionless
variable if we wish to see the attractor behaviour which is of central interest here.

\mysection{The hydrodynamic attractor in hydrodynamics} Hydrodynamic theories
are described by sets of nonlinear partial differential equations. The key
simplification brought by the assumption of Bjorken flow is that the equations
of hydrodynamics reduce to ordinary differential equations. For example, the
evolution equation for the pressure anisotropy in conformal BRSSS theory
reads~\cite{Heller:2015dha,Florkowski:2017olj}
\bel{feqn}
C_{\tpi}\left(1 + \f{\pa}{12}\right) \pa' + \left(\f{C_{\tpi}}{3 w} +\f{ C_{\lambda_{1}}}{ 8 C_{\eta}}
\right) \pa^2 = \f{3}{2} \left(\f{8 C_\eta}{w} - \pa\right)
\ee
where the prime denotes a derivative with respect to $w$, and the
dimensionless constants $C_{\eta}, C_{\tpi}, C_{\lambda_{1}}$ are transport
coefficients (whose values in the case of \symm\ are known, see
e.g. \rfc{Florkowski:2017olj}). This equation is nonlinear, but it can be
solved in powers of $1/w$: this is the hydrodynamic gradient expansion whose
leading term reproduces the prediction of Navier-Stokes hydrodynamics. It also
posesses an attractor, which can be determined numerically by setting initial
conditions appropriately~\cite{Heller:2015dha}. It is important to observe
that the attractor becomes indistinguishable from the first order truncation
of the gradient series only for $w>0.7$. For smaller values of $w$, the
numerical solutions clearly decay to the attractor, not to the truncated
gradient series.

The pressure anisotropy in HJSW theory satisfies a second order nonlinear ordinary differential equation,
whose exact form can be found in Refs.~\cite{Aniceto:2015mto,Florkowski:2017olj}, and a similar analysis leads to
the numerical determination of its attractor solution (to which we shall return shortly). The point we wish
to make at this juncture is that we cannot proceed in the same way in \symm, because there we cannot write
down a closed differential equation like \rf{feqn}. To find the attractor in this case one has to find
another way. The approach explored in this Letter is to sum the hydrodynamic gradient expansion, whose leading
$240$ coefficients were obtained using the AdS/CFT correspondence in \rfc{Heller:2013fn}. In the following we
discuss the properties of the series and the summation, using HJSW theory as a testing ground.

\mysection{Large order behaviour} In any conformal theory the general form of
the gradient expansion of the pressure anisotropy for Bjorken flow is (see,
e.g.~\rfc{Florkowski:2017olj})
\bel{rgradex}
\pa(w) = \sum_{n=1}^{\infty} \pac_n \, w^{-n}.
\ee
The coefficients $\pac_n$ have been calculated to high order in \symm~\cite{Heller:2013fn}, in kinetic
theory~\cite{Heller:2016rtz,Florkowski:2017olj}, as well as in various hydrodynamic
theories~\cite{Heller:2015dha,Aniceto:2015mto}. It is now well established that this series has a vanishing
radius of convergence. In many cases, at large order $n$ the coefficients grow in a way consistent with the
Lipatov form~\cite{Lipatov:1976ny}
\bel{lipatov}
\pac_n \sim \f{n!}{A^n},
\ee
where $A$ is a real parameter. This formula implies linear behaviour of the
ratio of neighbouring coefficients $\pac_{n+1}/\pac_n \sim n/A$.  For the case
of the gradient series in BRSSS hydrodynamics this linear behaviour can be
seen in the left-hand plot of \rff{fig:diverge}. In the case of HJSW theory
however the pattern is much more complex, as seen in the right-hand plot.  The
reason for this is that HJSW hydrodynamics instead of a single, purely damped
nonhydrodynamic mode has a pair of modes with complex conjugate
frequencies~\cite{Aniceto:2015mto}.  Furthermore, this signals that the
hydrodynamic gradient expansion in this case is an element of a two-parameter
transseries~\cite{Aniceto:2011nu} (while in BRSSS hydrodynamics the
transseries involves only one parameter).

The analysis of \rfc{Aniceto:2015mto} can be used to find the following
approximate formula describing the leading large $n$ behaviour
\bel{newlip}
\pac_{n}\sim \f{n!}{A^{n}} \cos\left((n+ \beta_R)\phi -\psi +
\beta_I\log\left(\f{A}{n + \beta_R}\right)\right)
\ee
for some real numbers $A, \phi, \psi, \beta_R, \beta_I$ (the $\psi$ appearing
above is the phase of the Stokes constant of the transseries). Due to the
oscillating factor this formula is not as useful as \rf{lipatov}, but it does
qualitatively capture the complex pattern in \rff{fig:diverge}.

Importantly, if one plots the ratio of coefficients of the gradient expansion
of \symm, calculated using the results in \rfc{Heller:2013fn}, one finds a
picture very similar to the right-hand plot of \rff{fig:diverge}.  This
happens because HJSW theory was constructed to reproduce the dominant
nonhydrodynamic modes of \symm, which results in the close similarity of the
large-order behaviour. This makes it a useful testbed for assessing the utility
of Borel summation in this context, as explained below.

\begin{figure}
\begin{center}
\includegraphics[height=0.3\textheight]{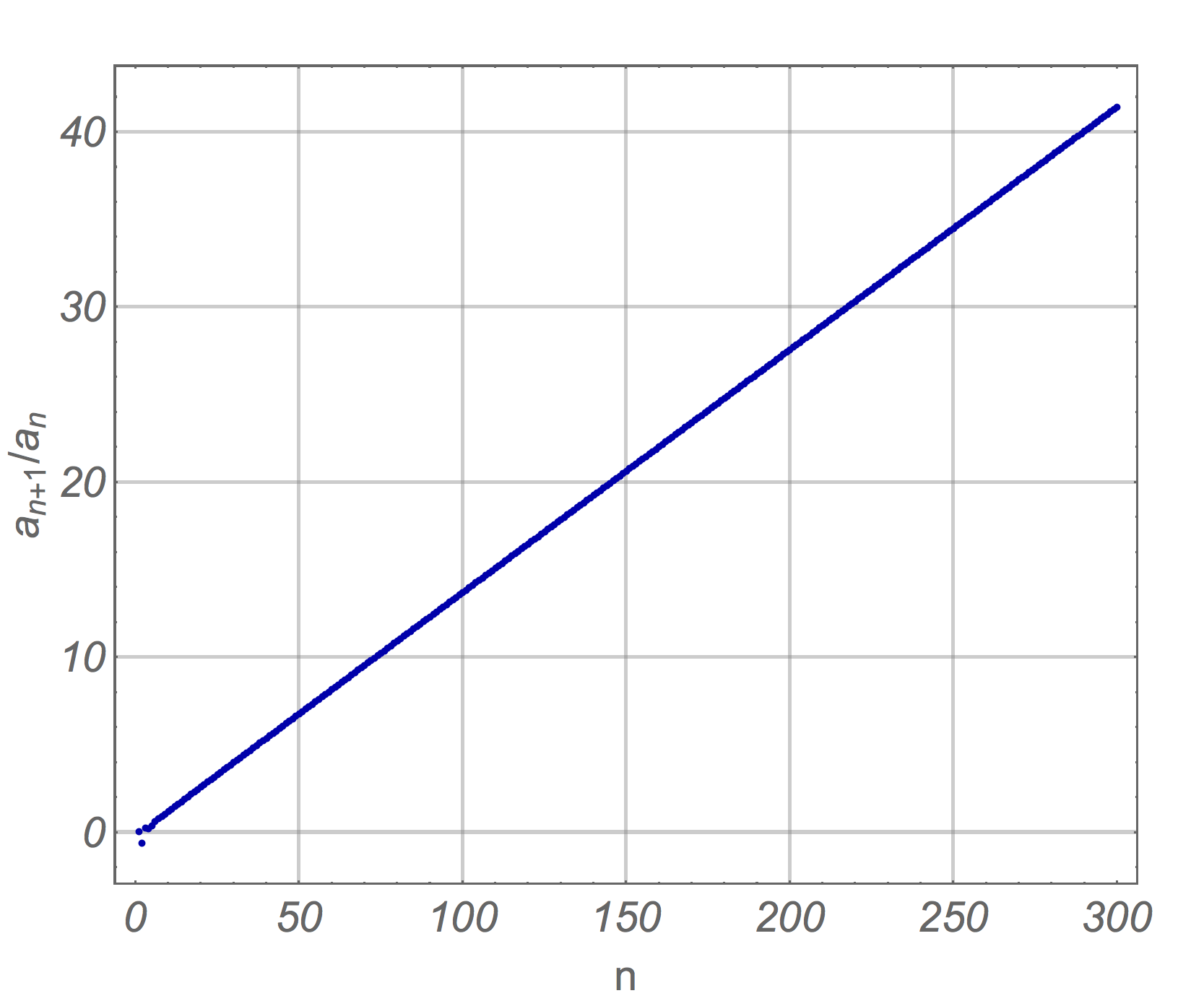}
\includegraphics[height=0.3\textheight]{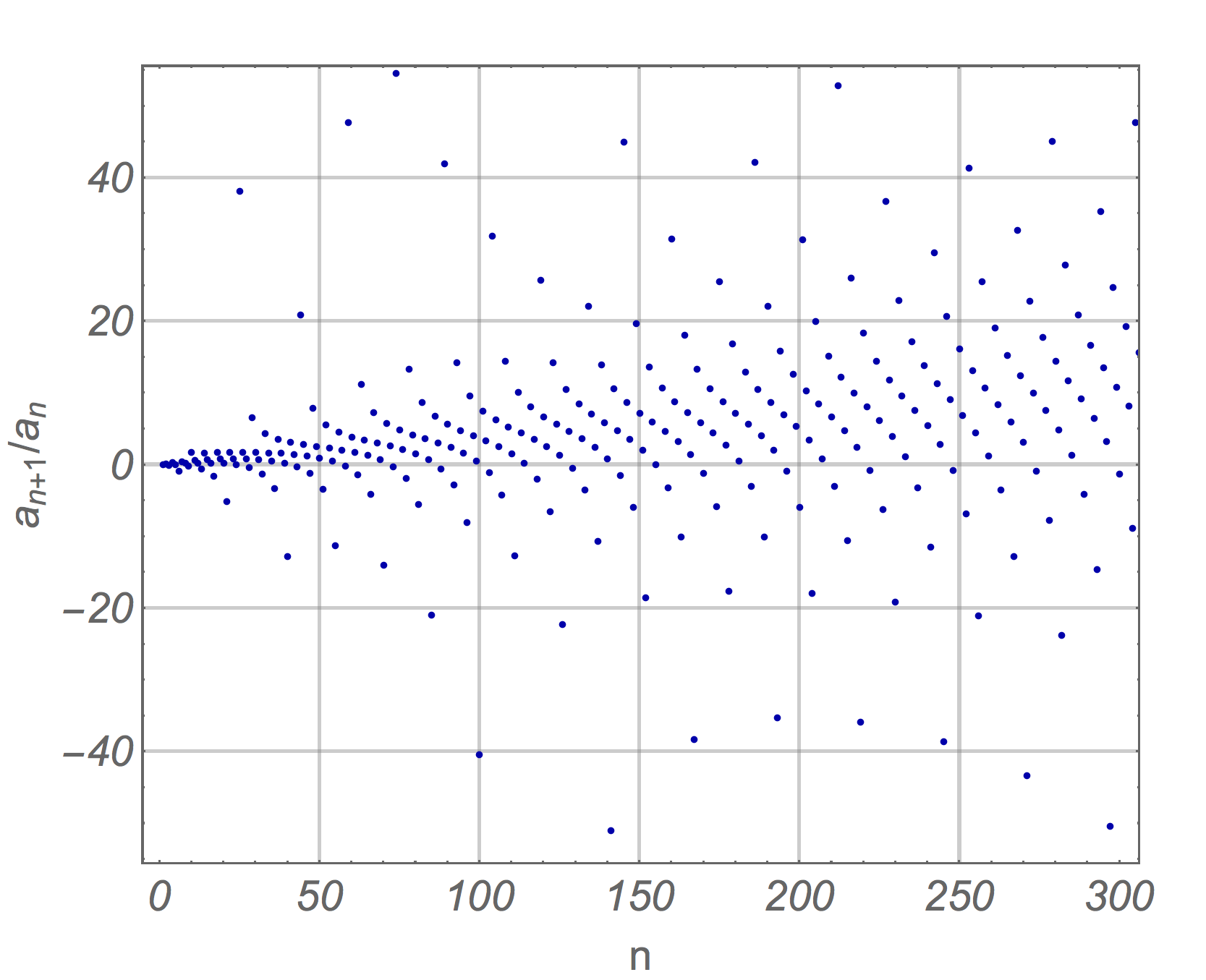}
\caption{
  The large order behaviour of the gradient series
  for BRSSS and HJSW hydrodynamics.}
\label{fig:diverge}
\end{center}
\end{figure}

\mysection{The attractor from Borel summation}
The Borel transform of the gradient series removes the dominant factorial growth of the expansion coefficients:
\bel{borel}
\bort \pa(\xi) = \sum_{n=1}^\infty \frac{\pac_n}{n!} \, \xi^{n}.
\ee
This series will typically define an analytic function within a disc around
the origin in the complex $\xi$ plane. The Borel sum of the series is defined
by the Laplace transform:
\bel{iborel}
\ibort \pa(w) = w \int_{\cal C} d\xi \, e^{- w \xi}\, \borta \pa(\xi),
\ee
where $\borta \pa$ is the analytic continuation of the Borel
transform \rfn{borel} and $\cal C$ is a contour connecting~$0$~and~$\infty$.

The analytic continuation of $\bort \pa(\xi)$ (performed using Pad\'e
approximants) necessarily contains singularities responsible for the vanishing
radius of convergence of the original series. The singularities appearing in
the cases of interest here have been discussed at length in the
literature. For BRSSS theory one finds a branch point on the real
axis~\cite{Heller:2015dha}, which introduces a complex ambiguity in the Borel
summation, given by the difference in the values obtained for \rf{iborel} by
integrating above and below the cut.  As stressed earlier, this complication
does not arise in the case of \symm~\cite{Heller:2013fn}, or HJSW
hydrodynamics~\cite{Aniceto:2015mto} where the branch points appear away from
the real axis. This means that one can perform the integral in \rf{iborel} by
integrating over real values of $\xi$ from zero to infinity.  In practice,
this integral has to be performed numerically for a set of values of $w$.

\begin{figure}
\begin{center}
\includegraphics[height=0.4\textheight]{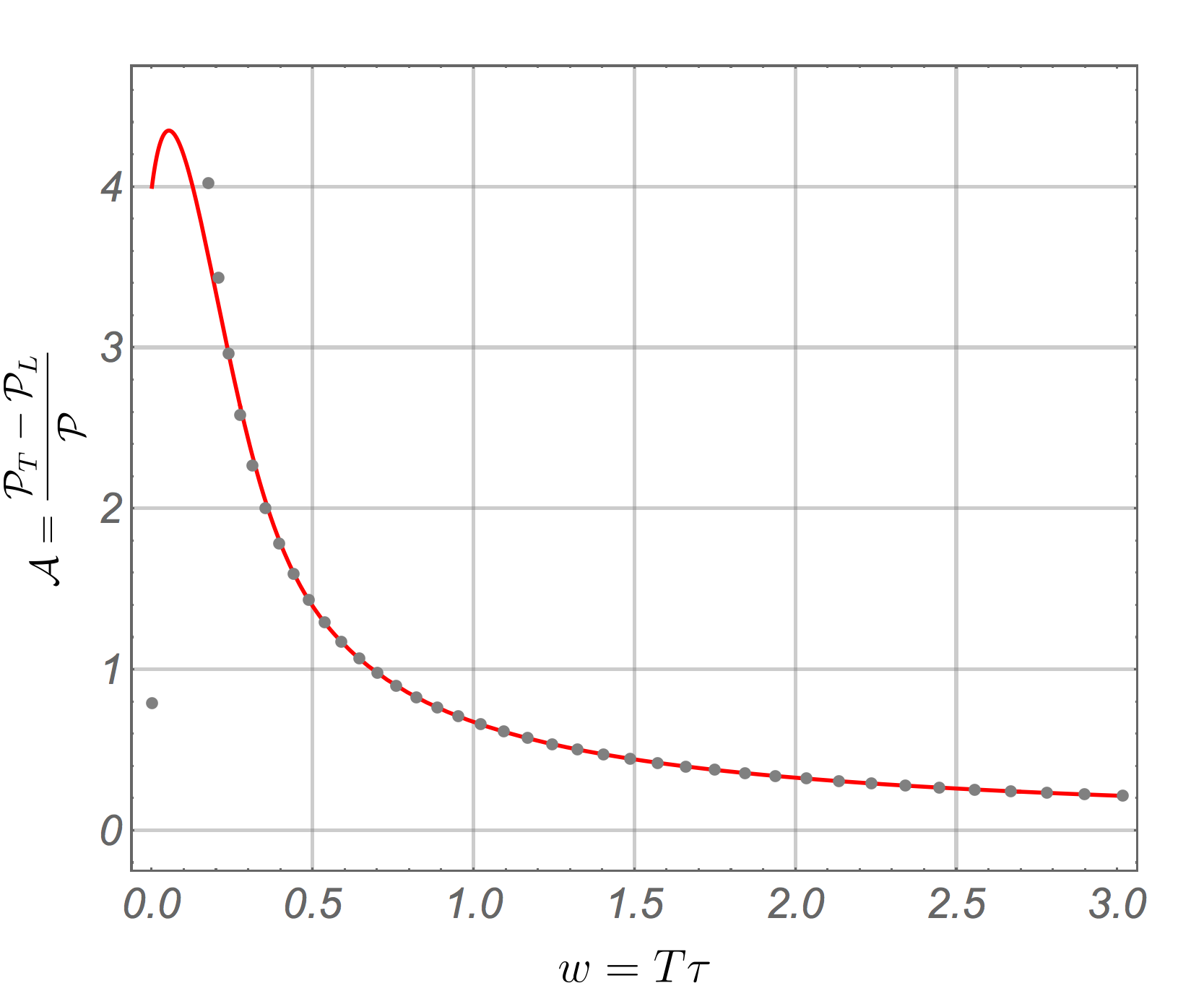}
\caption{The attractor in HJSW theory, calculated numerically from the
  hydrodynamic equations (red curve), compared with the results of Borel
  summation (gray dots).} 
\label{fig:hjsw}
\end{center}
\end{figure}

To gauge the effectiveness of this method we begin with HJSW theory, which
from this perspective offers qualitatively the same kind of challenge as
\symm. The series can be calculated numerically to essentially arbitrarily
high order~\cite{Aniceto:2015mto}, but here we will use only the first $240$
terms, to have a fair testing ground for the \symm\ case, where the cost of
calculating the coefficients is much higher, and at this time only $240$ are
available~\cite{Heller:2013fn}.
As is clear from \rff{fig:hjsw}, the Borel summation tracks the numerically determined attractor
closely down to $w\approx 0.4$, and is still quite reasonable at $w\approx
0.3$.

\mysection{The attractor of \symm}
The procedure described above can readily be applied to the hydrodynamic
gradient expansion of \symm\ using the results of \rfc{Heller:2013fn}, where
the expansion coefficients of the energy density in powers of $\tau^{-2/3}$
were calculated up to order $240$. Using equations \rfn{PLT} and \rfn{rdef},
these results can be translated into coefficients of the pressure
anisotropy \rfn{rgradex}.  As mentioned earlier, their ratios qualitatively
follow the pattern described by the approximate formula \rfn{newlip} (and seen
in the lower plot in \rff{fig:diverge}). These coefficients can be used to
calculate the Borel transform and its analytic continuation (using a diagonal
Pade approximant), which one can integrate numerically for a range of values
of $w$ exactly as described above for the case of HJSW theory. The
result is reproduced quite accurately by the rational function
\bel{apad}
\pa_0 = \frac{2530 w-276}{3975 w^2-570 w+120}
\ee
for essentially all values of $w>0$.

\begin{figure}
  \begin{center}
 \includegraphics[height=0.4\textheight]{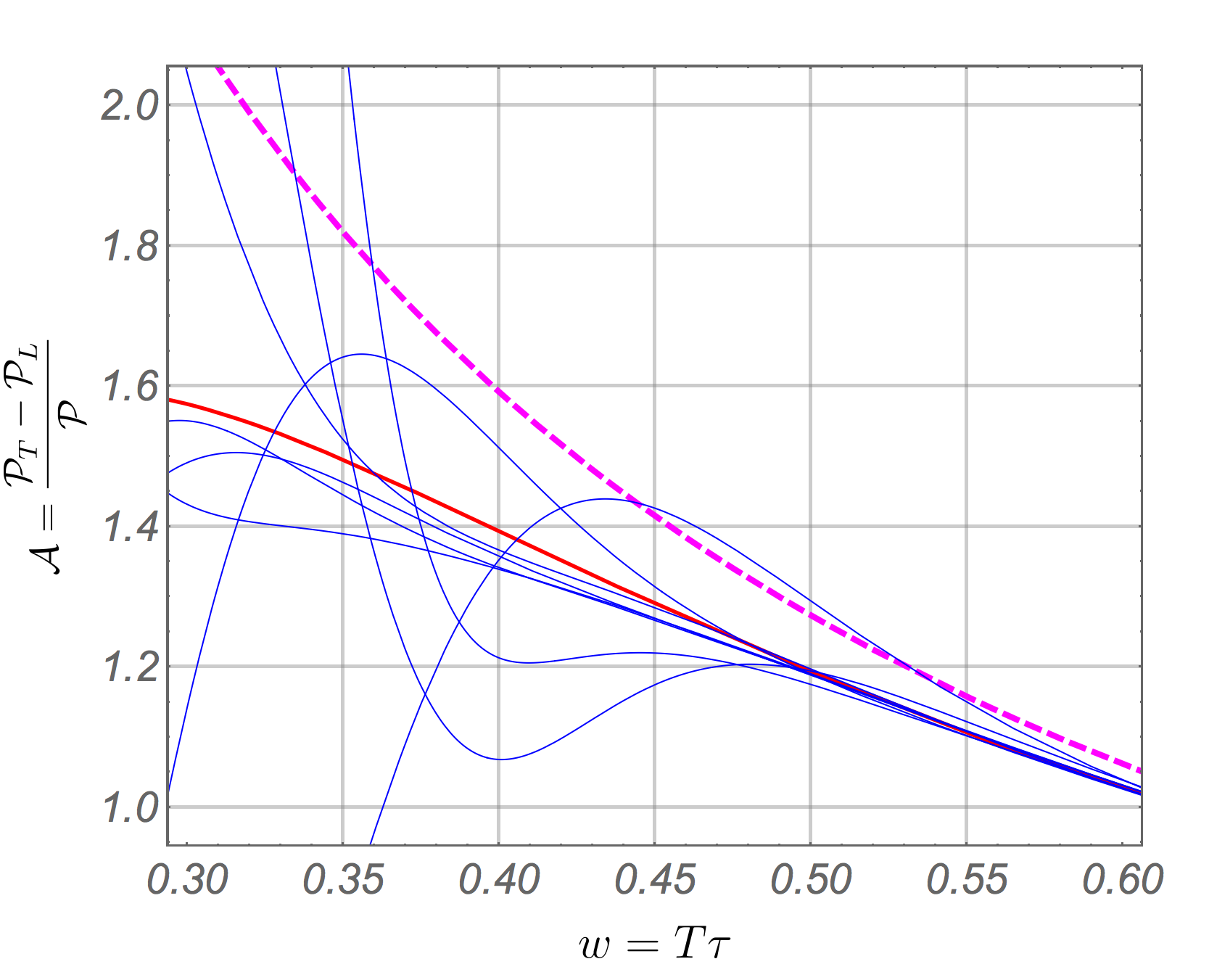}
\caption{The attractor of \symm\ plasma (the red curve), along with results of
  numerical simulations using AdS/CFT (blue curves); the dashed magenta curve
  represents the hydrodynamic gradient expansion truncated at first order.  }
\label{fig:attrSYM}
\end{center}
\end{figure}

The attractor determined by Borel summation as described above can be expected
to reliable down to $w\approx 0.3$, but its utility is best judged by
comparing it with the results of numerical holography simulations of Bjorken
flow. A large number of such flow histories was studied in
\rfc{Jankowski:2014lna} following the earlier work of
Refs.~\cite{Heller:2011ju,Heller:2012je}.
The distinguished role played by the
attractor is most prominently visible by considering values of $w$ for which
it differs appreciably from the truncated gradient expansion.  A selection of
solutions which reach the attractor at such early times is plotted in
\rff{fig:attrSYM}, along with the rational fit $\pa_0$ given in \rf{apad}.  We
see there the same kind of striking behaviour as seen at the level of
hydrodynamics in \rfc{Heller:2015dha}.

In \rfc{Jankowski:2014lna} the transition to hydrodynamics was defined in
terms of the pressure anisotropy matching the truncated gradient expansion
using the third order result from \rfc{Booth:2009ct}.  Each numerical solution
followed the hydrodynamic prediction at sufficiently late times. The threshold
was found to lie in a range of values of $w$ centered around $w_H=0.65$, with
a large pressure anisotropy $\pa(w_H)\approx 0.7$. However, given the results
presented here it is tempting to think of hydronization in terms of reaching
the attractor, which implies an even smaller value of $w_H$ and a
correspondingly higher value of $\pa(w_H)$. For most of the histories shown in
\rff{fig:attrSYM} the pressure anisotropy at hydrodynization exceeds 100\%,
showing that this observable can exhibit universal behaviour even in the
highly nonequilibrium regime. It should however be stressed that this is not a
claim concerning the behaviour of generic histories of the flow: the hydronization time depends
on the initial conditions.

The calculation described above leaves a number of open problems. One of them
is the explicit computation of leading transseries coefficients in the lowest
instanton sectors. Such a calculation would make it possible to extend the
range where the summation can be trusted to lower values of $w$. It would also
allow us to verify that the resurgence relations written down in
\cite{Aniceto:2015mto} connecting coefficients of different transseries
sectors are satisfied. Another interesting problem to pursue would be a
calculation of the \symm\ attractor directly in the holographic
representation. An attempt of this kind was recently made by
Romatschke~\cite{Romatschke:2017vte}, who tried to find the special initial
condition corresponding to the attractor solution, which he was then able to
estimate by evolution using the bulk Einstein equations. The results presented
in that paper are in good agreement with those presented here for $w>0.4$,
while at lower values of $w$ the method of \rfc{Romatschke:2017vte} suggests
that the true attractor may flatten out around $w=0.4$ and for smaller values of
$w$ lies somewhat below what is seen in \rff{fig:attrSYM}.

\mysection{Conclusions} The goal of hydrodynamics is to mimic universal,
late-time behaviour of systems tending toward
equilibrium~\cite{Florkowski:2017olj}. The BRSSS
philosophy~\cite{Baier:2007ix}, which can be seen as an incarnation of the
effective field theory paradigm, tells us to match leading terms of the
gradient expansion of hydrodynamics with the corresponding terms calculated in
the underlying microscopic theory. The developments of the past couple of
years suggest that it could make sense to set a more ambitious goal: to try to
reproduce the attractor of the underlying theory at the level of
hydrodynamics. While the attractor matches low orders of the gradient series
at sufficiently late times, earlier on it is different, and the difference
depends on the parameters of the theory. Even for \symm\ the attractor is
quite distinct from first (or second order) hydrodynamics. For QCD plasma,
which almost certainly has a larger relaxation time, this distinction should
be even more pronounced. This could have important consequences for the
interpretation of observables sensitive to early time dynamics.


At this time it would be most useful and
interesting to find tractable examples which relax some of the technical
elements which we have relied on, such as boost invariance and conformal
symmetry. An important point will be to understand which observables reveal
attractor behaviour. Any progress should be of interest not only in the
context of quark-gluon plasma, but also for other areas of
physics~\cite{Brewer:2015ipa,Bluhm:2017rnf}.

\ackno
I would like to thank J.~Casalderrey-Solana, M.P.~Heller and P.~Romatschke for 
useful discussions. This work was supported by the National Science
Centre Grant No. 2015/19/B/ST2/02824. 

\bibliographystyle{utphys}
\bibliography{seebib}{}

\end{document}